\documentclass{article}

\usepackage{spconf,amsmath,graphicx}
\usepackage{tikz,pgfplots}

\usepackage{algorithm}
\usepackage{algorithmic}
\usepackage[export]{adjustbox}
\usepackage{pgfplotstable}
\usepackage{booktabs}
\usepackage{amssymb}
\usepackage{multirow}
\usepackage{booktabs}
\usepackage{array}
\usepackage{colortbl}
\usepackage{csvsimple}
\usepackage{makecell}

\usetikzlibrary{decorations}
\usetikzlibrary{decorations.markings}
\usetikzlibrary{plotmarks}
\usetikzlibrary{positioning} 
\pgfplotsset{compat=1.13}
\usepgfplotslibrary{groupplots}
\usetikzlibrary{matrix,spy}

\def\x{{\mathbf x}}

\newcommand{\eg}{\emph{e.g.}}
\newcommand{\ie}{\emph{i.e.}}
\newcommand{\etc}{\emph{etc}}

\newcommand{\Kq}[1]{K_{#1}}
\newcommand{\T}[1]{R_{#1}}

\title{Accelerating the Registration of Image Sequences by \\
	Spatio-temporal Multilevel Strategies}
\name{Hari Om Aggrawal$^{1}$ \qquad Jan Modersitzki$^{1,2}$ \thanks{The financial support by the Federal Ministry of Education and Research of Germany in the framework of MED4D (project number 05M16FLA)}}
\address{$^{1}$ Institute of Mathematics and Image Computing, University of L\"ubeck, Germany \\ $^{2}$Fraunhofer Institute for Digital Medicine MEVIS, L\"ubeck, Germany}

\begin{document}
%
\maketitle
%

\begin{abstract}

Multilevel strategies are an integral part of many image registration algorithms. 
These strategies are very well-known for avoiding undesirable local minima, providing an outstanding initial guess, and reducing overall computation time. 
State-of-the-art multilevel strategies build a hierarchy of discretization in the spatial dimensions. 
In this paper, we present a spatio-temporal strategy, where we introduce a hierarchical discretization in the temporal dimension at each spatial level. 
This strategy is suitable for a motion estimation problem where the motion is assumed smooth over time. 
Our strategy exploits the temporal smoothness among image frames by following a predictor-corrector approach. 
The strategy predicts the motion by a novel interpolation method and later corrects it by registration. 
The prediction step provides a good initial guess for the correction step, hence reduces the overall computational time for registration. 
The acceleration is achieved by a factor of 2.5 on average, over the state-of-the-art multilevel methods on three examined optical coherence tomography datasets.

 
\end{abstract}

\begin{keywords}
	Groupwise image registration, spatio-temporal, multilevel, acceleration, motion estimation
\end{keywords}
\section{Introduction}
\label{sec:intro}

%

Motion estimation is a preliminary step in many medical imaging applications, \eg, for motion correction \cite{Kraus2015}, motion modeling \cite{McClelland2013,DeCraene2012}, \etc. 
Generally, we quantify the motion in terms of displacement fields and estimate them by solving an image registration problem \cite{McClelland2013}. 
Typically, registration models are highly non-convex in nature \cite{2009-FAIR}. 
Therefore, a good initial guess becomes essential to avoid undesirable local minima. 
Moreover, registration is an ill-posed problem that requires a good regularization strategy to solve the problem and obtain a desirable, \eg, a locally smooth displacement field \cite{2009-FAIR}.

Multilevel methods not only provide a good initial guess and an implicit regularization but
also accelerate the registration \cite{2009-FAIR}. 
State-of-the-art multilevel methods smooth the given images and discretize them at different spatial resolutions. 
Image smoothing leads to the objective function smoothing and smooths out many undesirable local minima.
Initial computations on a coarse resolution are computationally cheap and provide a good initial guess for the optimization problem at a finer resolution. 
Hence, this reduces the number of iterations and accelerates the registration. 
This strategy is also applicable for a spatio-temporal data. 
However, a multilevel discretization only in spatial dimensions does not exploit scale-space information in the spatio-temporal space \cite{Kai2019, 2009-FAIR}.

This paper motivates hierarchical discretization both in space and in time. 
The proposed strategy is suitable for motion estimation problems where the movement of an object over time is assumed to be small for a short interval of time and displacement fields are considered temporally smooth \cite{DeCraene2012}. 

The existing methods exploit the temporal smoothness either by parameterizing displacement fields in terms of temporally smooth basis functions \cite{DeCraene2012} or by introducing an explicit Tikhonov regularizer \cite{Kraus2015}. 
These approaches require the tuning of the number of basis functions or a regularization parameter to achieve the desired result. 
Generally, this demands to solve the optimization problem many times, which is a computationally expensive task and not suitable for live-imaging applications \cite{Kolb2019}.

Our approach exploits the temporal smoothness through the spatio-temporal multilevel discretization. We assume the temporal smoothness implies that the neighboring frames are highly correlated. 
This means a good approximation of displacement field for a frame is possible through the interpolation of displacement fields of the neighboring frames. 
This reasoning inspired us to define the spatio-temporal multilevel strategy based on a predictor-corrector approach. 

Here, we perform registration with a few of the images from the given image sequence and predict displacements for the rest of the images by a novel interpolation method. 
Afterward, we correct the prediction by registration along with the rest of the images. 
We assume that the approximation from the prediction step serves as a good initial guess for registration in the correction step. 
Hence, it reduces the overall computational time. 
We achieved acceleration by a factor of 2.5 on average, over the existing multilevel strategies.

In the forthcoming sections, we introduce an image registration framework and discuss the proposed spatio-temporal multilevel strategy and a novel interpolation method. We also demonstrate the effectiveness of the proposed strategy on three real datasets and conclude our findings.

\begin{section}{Image Registration Model}
	
Our goal is to register images $\T{i}$ of any spatial dimension acquired at $n$ timepoints. We assume that a groupwise registration scheme (GIR) is available that allows to register any subset $\Kq{} \subset \Kq{T}:={1,\ldots,n}$ of images from the sequence, that is to compute optimal transformations $Y=(y_k,k\in \Kq{})$. For this paper, we use the framework outlined in \cite{2009-FAIR}. However, the spatio-temporal multilevel approach can be used for any other meaningful registration strategy. 
With this, the goal of GIR is to determine a minimizer of 
 \begin{equation}\label{regmodel}
 	J(Y)
	:=\sum_{i\in \Kq{}}\sum_{j\in \Kq{}} D(\T{i}\circ y_i,\T{j}\circ y_j)
	+S(Y)+P(Y),
 \end{equation}
where $\T{i}\circ y_i$ denotes the transformed image,
$D$ measure dissimilarity of images, $S$ is a regularization and $P$ is a penalty; see \cite{2009-FAIR} details. For the distance, we use a recently proposed correlation based approach, cf. \cite{Kai2019}. The regularization is relevant in this paper. However, the penalty is added to prevent for global distortions \cite{Bhatia2004},
\[
	P(Y)=\lambda\textstyle\|1/|\Kq{}|\sum_{j\in \Kq{}} y_j - \mathrm{Id}\|^2,
	\quad
	\mathrm{Id}\mbox{ is the identity,}
\] 
where $\lambda$ is a regularization parameter; see Sec.~\ref{sec:results} for choices.
 
Of course, any numerical optimization scheme can be used; here we used the L-BFGS outlined in \cite{2009-FAIR}.

\end{section}

\newcommand{\GIR}[1]{\mathrm{GIR}(#1)}
\newcommand{\xx}[2]{#1(#2)}
\newcommand{\ym}[1]{\xx{y}{#1}}
\newcommand{\zm}[1]{\xx{z}{#1}}
\newcommand{\all}{\mathrm{all}}

\begin{section}{Spatio-temporal Multilevel Strategy}
	
State-of-the-art multilevel methods generally build a hierarchy of discretizations in the spatial dimensions and perform registrations sequentially from coarse to fine level; for details see, e.g., \cite[ch.~6]{2009-FAIR}. In this paper, we present a combined approach, where an additional temporal multilevel strategy is added at each spatial level. The purpose is to provide an outstanding starting point for the final GIR such that the optimization can be performed much faster and much more robust; see also the result section.
	
The spatio-temporal multilevel strategy has two components. In a first step, data will be computed on different spatial discretization levels $\ell$ ranging from coarse to fine. In a second step, a simple binary tree is used to define an additional temporal structure $\Kq{q}$, $q=0,\ldots,T$. Here, the finest temporal set $\Kq{T}$ contains all $n$ images, $\Kq{T-1}$ every second, 
$\Kq{T-2}$ every fourth, and so on, until $\Kq{0}$ which contains only the left, middle, and right images in time; see also Fig.~\ref{fig:stml}. The procedure can obviously be adapted to any number of given images.

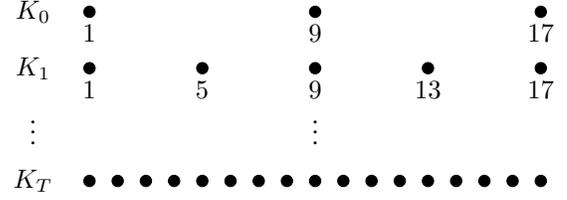
\begin{figure}[t]
\centering

\begin{tikzpicture}[scale=0.75]
	\draw(-1,-0.5) node {$\Kq{0}$};
	\foreach \x/\t in {0/1,4/{9},8/{17}}{
		\draw[black,fill](\x,-0.5) circle (0.1)
			node [below=0.5mm] {$\t$}
			;
	}

	\draw(-1,-1.5) node {$\Kq{1}$};
	\foreach \x/\t in {0/1,2/{5},4/{9},6/{13},8/{17}}{
		\draw[black,fill](\x,-1.5) circle (0.1)
			node [below=0.5mm] {$\t$}
			;
	}
	\draw(-1,-2.5) node{$\vdots$};
	\draw(4,-2.5)  node{$\vdots$};

	\draw(-1,-3.5) node {$\Kq{T}$};
	\foreach \x in {0,...,16}{
		\draw[black,fill]({\x/2},-3.5) circle (0.1)
			;
	}
\end{tikzpicture}

	\caption{Illustration of a binary tree based discretization scheme in the temporal dimension for $n = 17$. The set $\Kq{0}$ has elements $\{1,9,17\}$, $\Kq{1}$ has $\{1,5,9,13,17\}$, and the last set $\Kq{T}$ contains $n$ elements $\{1,\dots,17\}$.}
	\label{fig:stml}
\end{figure}

The idea is to exploit the spatial and temporal relation between the images. On each spatial level $\ell$, a predictor-corrector approach is introduced, which is based on different subsets $\Kq{q}$ of the temporal sequence.


For ease of presentation, we denote the image presentation at level $\ell$ and time $t_k$ by $\T{}(\ell,k)$ and the corresponding transformation which are to be computed by $\ym{\ell,q,\all}$
if all images are considered and by $\ym{\ell,q}$ if only transformations belonging to $\Kq{q}$ are considered. 
Note that the transformations depend on the spatial level as well as on the subset $\Kq{q}$ of images taken into account.

We start by outlining the procedure for the spatially coarse presentation of the images $\ell=\mathrm{coarse}$, which differs from the following finer levels. 
For the first temporal level $q=0$, we solve $\GIR{\ell,q}$ with starting values $\zm{\ell,q} := \mathrm{Id}$ and obtain optimal transformation $\ym{\ell,q}$, 
i.e. optimal with respect to the subset $\Kq{q}$.

On the next temporal level $q + 1$, starting values $\zm{\ell,q+1}$ are predicted as injections 
of the estimates $\ym{\ell,q}$ and linear interpolation for the intermediates;
for $k\in\Kq{q+1}\backslash \Kq{q}$,
\[
  \zm{\ell,q+1,k}
  =0.5\ym{\ell,q,\mathrm{left}(k)}+0.5\ym{\ell,q,\mathrm{right}(k)},
\]
where $\mathrm{left}(k)=\max\{z<k,z\in \Kq{q}\}$ and
$\mathrm{right}(k)=\min\{z>k,z\in \Kq{q}\}$, respectively.

In a correction step, we obtain $\ym{\ell,q+1}$ from $\GIR{\ell,q+1}$ 
using the predicted transformations $\zm{\ell,q+1}$ 
as an educated starting guess.
This procedure is repeated until the finest temporal level $T$ is reached.

After finishing the registration for the coarse spatial level $(\ell-1)$ 
for all timepoints, we continue with the finer spatial level $\ell$.
In contrast to the coarse level, we now also have estimates 
$\ym{\ell-1,T,\all}$ from the previous coarse level for prediction at the finer level.
We incorporate this knowledge into our predictor scheme and present the novel interpolation strategy \eqref{eq:ls} which is well suited for this particular application.

For the first temporal level $q=0$, we use $\ym{\ell-1,T}$ as an initial estimates 
$\zm{\ell,0}$ and compute $\ym{\ell,0}$ from $\GIR{\ell,0}$.



On the next temporal level, we construct the starting guess 
by enforcing proximity to the solution of the previous temporal level 
as well as to the local changes over time. These local changes are estimated
from the previous temporal level, or from the results of the last spatial level, depending on the current temporal level $q$.


To be more formal, our starting guesses $\zm{\ell,q+1,\all}$ are the minimizers of 
the simple least squares problems, which are solved for all components independently.
We denote the $i$-th components of the estimator, solution of the previous temporal level, and a reference for local changes
by $\zeta_k=\zm{\ell,q+1,k}_i$, $\eta_k=\ym{\ell,q,k}_i$, 
and $\omega_{k}:=\zm{\ell,q,k}_i$ for $q>0$ and 
$\omega_{k}:=\ym{\ell-1,T,k}_i$ for $q=0$, respectively. 

The estimator $\zeta_k$ is determined as the minimizer of
\begin{equation}\label{eq:ls}
	\sum_{k\in\Kq{q}} (\zeta_k - \eta_k)^2
	+\beta
	\sum_{k=0}^{n-1}\Big[(\zeta_{k+1}-\zeta_k) - (\omega_{k+1}-\omega_{k}))\Big]^2;
\end{equation}
see also Fig.~\ref{fig:interpolation} for an illustration,
and Sec.~\ref{sec:results} for choices of~$\beta$.

The estimates $\zm{\ell,q+1}$ are used for $\GIR{\ell,q+1}$ to compute $\ym{\ell,q+1}$ and the
procedure is continued until $q=T$, i.e $\ym{\ell,T,\all}$ is obtained on level $\ell$, 
and further it continues until $\ell$ reaches the finest spatial level.

Note that each individual least squares problems \eqref{eq:ls} results in a $n$-by-$n$ tridiagonal matrix, which can be solved by Thomas'-Algorithm in just ${\cal O}(n)$; cf.~\cite{Higham2002}.


\begin{figure}[t]
	\centering
	\begin{minipage}[b]{\linewidth}
		
		\begin{tikzpicture}[
		every node/.style={inner sep=0pt,outer sep=0pt}]
		
		\begin{axis}[
		yscale = 0.8, xscale = 1,
		xlabel={$k\in \Kq{T}$ (ticks) and $j\in \Kq{q}$ (circles)}, 
		ylabel={}, 
		x label style = {yshift = -0.3cm},
		y label style = {yshift = 0.3cm},
		axis lines=left, ytick=\empty, 
		minor xtick = {0,0.02,...,1},
		ymin = 7,
		xmin = 0, xmax = 1.1,
		xticklabels = {};
		x tick label style = {yshift = -0.1cm},
		legend style={at={(axis cs:0.1,18)},anchor=south west}]
		\addplot[thin,dashed] 
		table[y index=1] {figures/prolongationNew.dat};			
		\addplot[only marks, mark=o, mark indices={1,2,3,4,5}, mark options={black},mark size=3pt] 
		table[y index=2] {figures/prolongationNew.dat};
		\addplot[thin,solid, red] 
		table[y index=3] {figures/prolongationNew.dat};
		\addplot[only marks, mark=o, mark indices={1,2,3,4,5}, mark options={black},mark size=3pt] 
		table[y index=4] {figures/prolongationNew.dat};

		\addlegendentry{$\omega_k$}
		\addlegendentry{$\eta_j$}
		\addlegendentry{$\zeta_k$}

		\end{axis}
		\end{tikzpicture}
	\end{minipage}
	\caption{%
		Novel interpolation strategy~\eqref{eq:ls} generates optimal starting guesses 
		$\zeta_k$ (solid line) for the temporal level $(q+1)$.
		Estimates $\omega_k$ (dashed line) 
		and registration results $\eta_j$ (circles) are supplied;
		note that $k\in \Kq{T}$ and $j\in \Kq{q}$.
	}
	\label{fig:interpolation}
\end{figure}
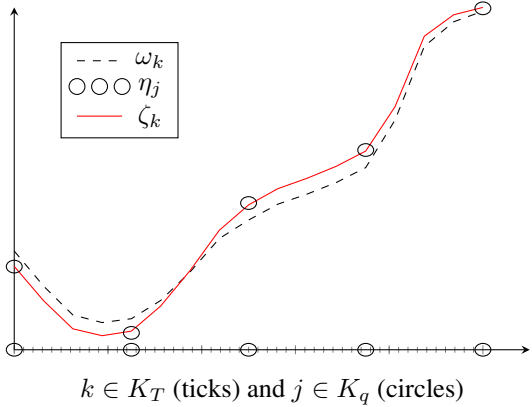

%
%
%
%
%
%
%
%
%
%
%
%
%
%
%
%

Note that from \eqref{eq:ls} we have estimates not only for $k\in \Kq{q}$ but for all $k\in \Kq{T}$. This enables us to stop the registration if a user supplied tolerance is already met in the $q$-th temporal resolution, i.e.
 \begin{equation}\label{stoppingrule}
     |D(\ym{\ell,q})-D(\ym{\ell,q-1})| \leq \epsilon,
 \end{equation}
 where $\epsilon$ is a user supplied tolerance. Note that this further reduces our computational time.

\end{section}


%
%
%
%
%
%
%
%
%
%

\section{Results}
\label{sec:results}

\begin{figure}[t]
	\centering
	\begin{minipage}[b]{0.3\linewidth}
		\includegraphics[width=0.9\columnwidth,height=1.5\columnwidth,frame]{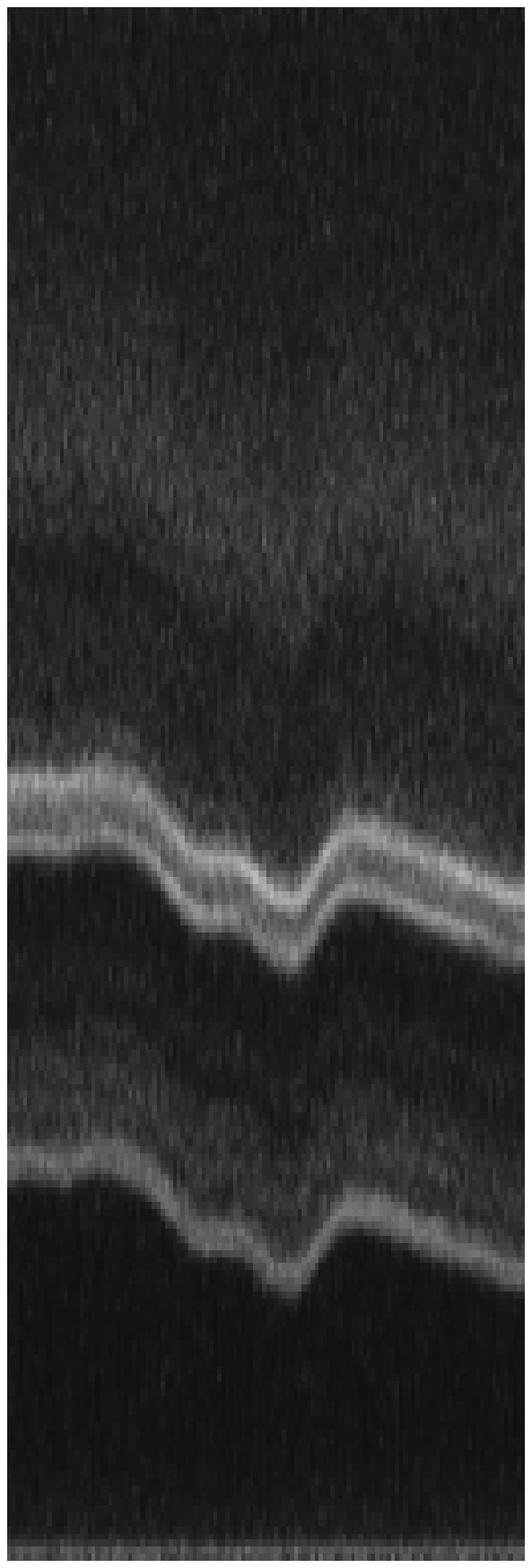}
		\centerline{\footnotesize{(a) Unregistered}}
	\end{minipage}
	\begin{minipage}[b]{0.3\linewidth}
		\includegraphics[width=0.9\columnwidth,height=1.5\columnwidth,frame]{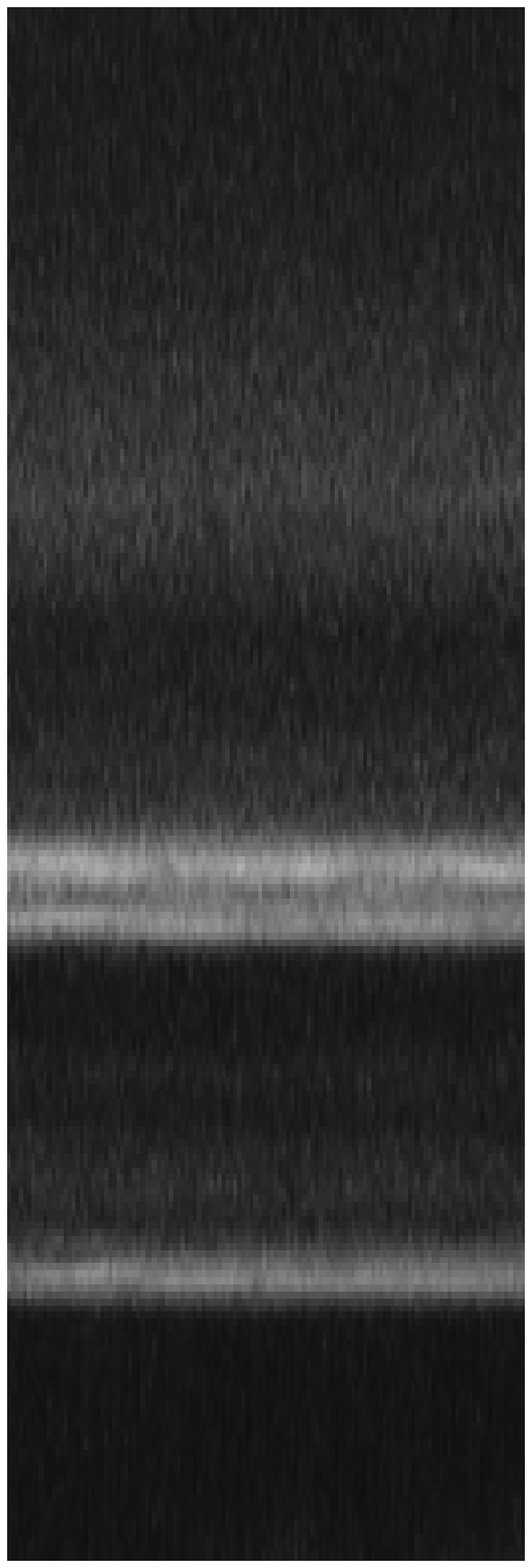}
		\centerline{\footnotesize{(b) SpML Method}}
	\end{minipage}  
	\begin{minipage}[b]{0.3\linewidth}
		\includegraphics[width=0.9\columnwidth,height=1.5\columnwidth,frame]{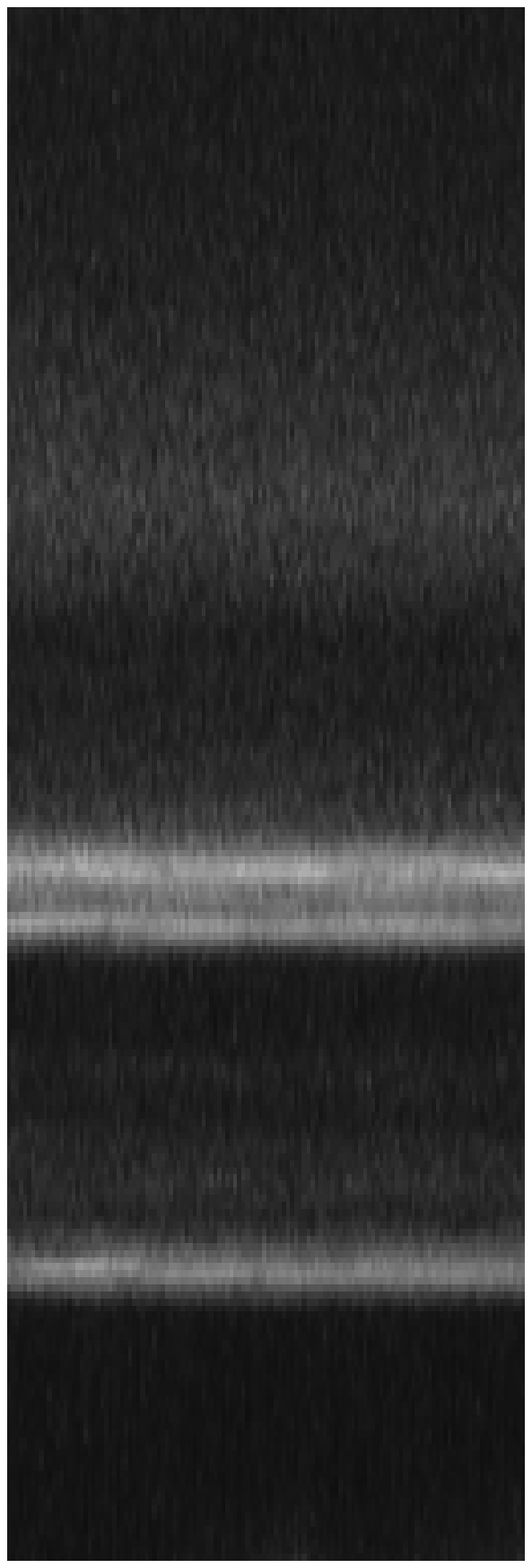}
		\centerline{\footnotesize{(c) STML Method}}
	\end{minipage}  
	\caption{
		The unregistered data (a) is transformed with estimates computed from GIR using SpML and STML method and results in (b) and (c) respectively. Here, we show the intensity variation over time only along a line in the 3D space. Horizontal and vertical directions represent time and space axis respectively. All images are at the same scale between [0,1].
		}
	\label{fig:lineprofile}
\end{figure}

\begin{table*}[ht]\label{table:OCTresults}
\centering
\caption{\label{table:OCTresults}Registration results for three 4D-OCT datasets}
\begin{tabular}{lrcrrrcrrr}\toprule
\multirow{2}{*}{\thead{Data (D) \\ Spatial Level  \\ (SL)}} &
\multirow{2}{*}{\thead{Spatial grid}} &
\multirow{2}{*}{\thead{Minimum of \\ $\rho_{i(i+1)}, i \in \Kq{T}$}} &
\multirow{2}{*}{\thead{$\lambda$}} &
\multicolumn{2}{c}{\thead{Reduction in D, in \%}} &
\multirow{2}{*}{\thead{Relative \\ difference in y \\ SpML vs STML, in \%}} &
\multicolumn{2}{c}{\thead{Run time (r) in sec.}} &
\multicolumn{1}{c}{\thead{Speedup}} \\
& & & & \thead{SpML} & \thead{STML} & & \thead{$r_s$, SpML} & \thead{$r_t$, STML} & \thead{$r_{s}/r_{t}$}\\
\midrule
\begin{filecontents*}{figures/OCTresults.csv}
dataName,ImgSize,rhoMin,alphavec,runtimeSp,runtimeSTU,runtimeRedSTU,reductSp,reductSTU,ydiff
D1 SL-1,29x10x10,0.972,100,42,23,1.8,1.7,0.9,2.6
D1 SL-2,57x20x20,0.939,100,51,16,3.1,1.5,1.0,4.7
D1 SL-3,113x40x40,0.915,100,447,154,2.9,2.9,2.7,5.3
D2 SL-1,57x20x10,0.996,1000,115,76,1.5,5.8,4.8,1.7
D2 SL-2,113x40x20,0.990,1000,490,174,2.8,10.4,9.3,1.1
D2 SL-3,225x80x40,0.970,1000,7077,2785,2.5,27.9,27.4,2.4
D3 SL-1,113x40x10,0.994,1000,145,73,2.0,3.1,2.4,1.8
D3 SL-2,225x80x20,0.980,1000,787,374,2.1,12.3,11.6,2.8
\end{filecontents*}
\csvreader[%
  head to column names,
  late after line=,
  table head=\hline,
  before line=
	\ifthenelse{\thecsvrow=4}
		{\\\midrule \multicolumn{7}{l}{D1 - total run time and speedup} & 540 & 193 & 2.8
		\\ \midrule}
		{
			\ifthenelse{\thecsvrow=7}
				{\\\midrule \multicolumn{7}{l}{D2 - total run time and speedup} & 7681 & 3035 & 2.5
				\\ \midrule}
				{\\}
	    },
  before first line=,
  after line = \ifthenelse{\thecsvrow=8}
				{\\\midrule \multicolumn{7}{l}{D3 - total run time and speedup} & 932 & 446 & 2.1}{},
  table foot = \\\midrule
]{figures/OCTresults.csv}{}%
{\dataName & \ImgSize & \rhoMin & \alphavec & \reductSp & \reductSTU & \ydiff &\runtimeSp & \runtimeSTU & \runtimeRedSTU}%
\\\bottomrule
\end{tabular}
\end{table*}

The acceleration achieved from the proposed spatio-temporal multilevel method is demonstrated on 4D optical coherence tomography (4D-OCT) datasets;
data courtesy of Robert Huber, Universit\"at zu L\"ubeck, Germany.
The data show scans of a posterior of eyes over time; see \cite{Kolb2019} for details.

We report speedup by a factor of 2.5 on average from the proposed spatio-temporal multilevel (STML) strategy over the state-of-the-art spatial multilevel (SpML) strategy \cite{2009-FAIR}.

Using \cite{2009-FAIR}, we compute the spatial multilevel representations as indicated in Tab.~\ref{table:OCTresults} for a total of $n=129$ timeframes. Note that we perform image smoothing only in the spatial dimensions. 
Tab.~\ref{table:OCTresults} reports results for three 4D-OCT datasets. 

We report a correlation coefficient \cite{2009-FAIR} between consecutive frames, \ie,  $\rho_{i(i+1)}, i \in \Kq{T}$ in Tab.~\ref{table:OCTresults}. 
For the examined datasets, $\rho_{i(i+1)}$ is close to the maximum possible value of $\rho$, \ie, 1. 
This indicates that the consecutive frames are highly correlated
and transformations between these frames are small and supposedly, temporally smooth over time. Fig.~\ref{fig:lineprofile}(a) clearly shows that the transformations are small and smooth over time. 


For the STML method, we start registration at the coarsest spatial level with $|\Kq{0}| = 17$ images out of the total $|\Kq{T}| = 129$ images. 
We linearly interpolate the estimated parameters to build a starting guess for GIR at the next temporal level $\Kq{1}$, where $|\Kq{1}| = 33$. 
These steps continue until the iterates satisfy the stopping criterion \eqref{stoppingrule}. 
At the next finer spatial level, we use estimates from the previous spatial level as a starting guess and register with $|\Kq{0}| = 17$ images. 
Now, we use \eqref{eq:ls} with $\beta=10^{-5}$ to build a starting guess at the next temporal level.
These steps continue until the finest spatial level.

We perform an affine-linear registration between frames which is enough for the alignment of structures in the examined datasets, see Fig.~\ref{fig:lineprofile} for results before and after registration. 
After registration, structures in the transformed images are perfectly aligned and the images are almost the same from both methods.

However, the STML method is faster by a factor of 2.5 on average, to the SpML method. 
Moreover, acceleration is observed at all spatial levels for all three datasets. 
The computation at the finest spatial level is the most contributing factor in the total run time, where we achieved even higher speed-ups than at a coarser spatial level; see the last column of Tab.~\ref{table:OCTresults}.

To compare results from the STML and the SpML method, we report two measures in Tab.~\ref{table:OCTresults}, \ie, reduction in dissimilarity measure $D$ from the unregistered state and the relative difference between estimates $y$ from the multilevel methods. 
For our experiments, the values of these measures roughly indicate that both methods converge in proximity as desired. 
This also shows that the STML strategy is robust concerning a multilevel discretization.

\section{Conclusion}
\label{sec:conclusion}

We proposed a spatio-temporal multilevel method to accelerate the registration of more than two images. We achieved acceleration by a factor of 2.5 on average, over the state-of-the-art multilevel method on the OCT dataset. The proposed predictor-corrector approach exploits the temporal smoothness intrinsic to dynamic image sequences without relying on any predefined temporal regularization. Moreover, with the novel interpolation method, the strategy implicitly provides a temporally smooth displacement field which is attractive for motion modeling applications \cite{DeCraene2012}. Furthermore, the accelerated registration would be beneficial for live medical imaging applications, \eg, surgical guidance \cite{Kolb2019}. Future work will address the application to dynamic image sequences from imaging modalities, \eg, PET, MRI.





\bibliographystyle{IEEEbib}
\bibliography{strings,refs}

\begin{thebibliography}{1}

\bibitem{Kraus2015}
Martin~F. Kraus and Joachim Hornegger,
\newblock {\em OCT Motion Correction}, pp. 459--476,
\newblock Springer International Publishing, Cham, 2015.

\bibitem{McClelland2013}
J.R. McClelland, D.J. Hawkes, T.~Schaeffter, and A.P. King,
\newblock ``Respiratory motion models: A review,''
\newblock {\em Medical Image Analysis}, vol. 17, no. 1, pp. 19--42, Jan. 2013.

\bibitem{DeCraene2012}
M.~De~Craene et.al.,
\newblock ``Temporal diffeomorphic free-form deformation: Application to motion
  and strain estimation from 3d echocardiography,''
\newblock {\em Medical Image Analysis}, vol. 16, no. 2, pp. 427--450, Feb.
  2012.

\bibitem{2009-FAIR}
J.~Modersitzki,
\newblock {\em {FAIR}: Flexible Algorithms for Image Registration},
\newblock SIAM, Philadelphia, 2009.

\bibitem{Kai2019}
K.~Brehmer, H.~O. Aggrawal, S.~Heldmann, and J.~Modersitzki,
\newblock ``Variational registration of multiple images with the {SVD} based
  {SqN} distance measure,''
\newblock in {\em Scale Space and Variational Methods in Computer Vision},
  Cham, 2019, pp. 251--262, Springer International Publishing.

\bibitem{Kolb2019}
J.~P. Kolb, W.~Draxinger, J.~Klee, T.~Pfeiffer, M.~Eibl, T.~Klein, W.~Wieser,
  and R.~Huber,
\newblock ``Live video rate volumetric {OCT} imaging of the retina with
  multi-{MHz} a-scan rates,''
\newblock {\em {PLOS} {ONE}}, vol. 14, no. 3, pp. 1--20, Mar. 2019.

\bibitem{Bhatia2004}
K.K. Bhatia, J.V. Hajnal, B.K. Puri, A.D. Edwards, and D.~Rueckert,
\newblock ``Consistent groupwise non-rigid registration for atlas
  construction,''
\newblock in {\em 2004 2nd {IEEE} ISBI: Macro to Nano}. {IEEE}.

\bibitem{Higham2002}
Nicholas~J. Higham,
\newblock {\em Accuracy and Stability of Numerical Algorithms},
\newblock SIAM, Jan. 2002.

\end{thebibliography}


\end{document}